\documentclass{appolb}
\usepackage{epsfig}

\begin{document}
\title{Mathisson Equations: Non-Oscillatory Solutions \\ in a
Schwarzschild Field%
}
\author{Roman Plyatsko, Oleksandr Stefanyshyn
\address Pidstryhach Institute for Applied Problems\\ in Mechanics and
Mathematics,\\ Ukrainian National Academy of Sciences\\ Naukova 3-b, 79060 Lviv, Ukraine}

\maketitle
\begin{abstract}
The Mathisson equations under the Frenkel-Mathisson supplementary condition are studied
in a Schwarzschild field. The choice of solutions, which describe the motions of the
proper center of mass of a spinning test particle, is discussed, and the calculation
procedure for highly relativistic motions is proposed. The very motions are important for
astrophysics while investigating possible effects of the gravitational spin-orbit
interaction on the particle's world line and trajectory.
\end{abstract}

\section{Introduction}
70 years ago Myron Mathisson has presented the equations describing the motions of a
spinning test particle in a gravitational field \cite{1}
\begin{equation}\label{1}
\frac D {ds} \left(mu^\lambda + u_\mu\frac {DS^{\lambda\mu}} {ds}\right)= -\frac {1} {2}
u^\pi S^{\rho\sigma} R^{\lambda}_{\pi\rho\sigma},
\end{equation}
\begin{equation}\label{2}
\frac {DS^{\mu\nu}} {ds} + u^\mu u_\sigma \frac {DS^{\nu\sigma}} {ds} - u^\nu u_\sigma
\frac {DS^{\mu\sigma}} {ds} = 0,
\end{equation}
where $u^\lambda$ is the 4-velocity of a spinning particle, $S^{\mu\nu}$ is the
antisymmetric tensor of spin, $m$ and $D/ds$ are, respectively, the mass and the
covariant derivative with respect to the proper time $s$; $R^{\lambda}_{\pi\rho\sigma}$
is the Riemann curvature tensor of the spacetime. (Throughout this paper we use units
$c=G=1$. Greek indices run 1, 2, 3, 4 and Latin indices 1, 2, 3; the signature of the
metric (--,--,--,+) is chosen.) Equations (\ref{1}), (\ref{2}) were supplemented by the
condition \cite{1}
\begin{equation}\label{3}
S^{\mu\nu} u_\nu = 0.
\end{equation}
(In special relativity condition (\ref{3}) was introduced by J.~Frenkel \cite{2}.) It is
known \cite{3, 4} that in the Minkowski spacetime the Mathisson equations
(\ref{1})--(\ref{3}) have, in addition to usual solutions describing the straight
worldlines, a family of solutions describing the oscillatory (helical) worldlines (as a
partial case, this family contains the circular solutions.)

In \cite{3} the oscillatory solutions of equations (\ref{1})--(\ref{3}) were connected
with {\it "Zitterbewegung"}. C.~M\"oller proposed another interpretation of these
solutions: he pointed out that 1) in relativity the position of the center of mass of a
rotating body depends on the frame of reference, and 2) condition (\ref{3}) is common for
the so-called proper and non-proper centers of mass \cite{5}. The usual solutions
describe the motion of the proper center of mass of a spinning body (particle), and the
helical solutions describe the motions of the family of the non-proper centers of mass
\cite{5}.

Later A. Papapetrou derived equations (\ref{1}), (\ref{2}) by the method which differs
from Mathisson's one \cite{6}, and, instead of (\ref{3}),  the non-covariant condition
\begin{equation}\label{4}
S^{i4} = 0
\end{equation}
was used \cite{7} in concrete calculations.

To avoid the superfluous solutions of equations (\ref{1}), (\ref{2}), W. Tulczyjew and
W.~Dixon introduced the covariant condition \cite{8}
\begin{equation}\label{5}
S^{\mu\nu} P_\nu = 0,
\end{equation}
where
\begin{equation}\label{6}
P^\nu = mu^\nu + u_\mu\frac {DS^{\nu\mu}}{ds}
\end{equation}
is the particle's 4-momentum. In contrast to relation (\ref{3}) the Tulczyjew-Dixon
condition (\ref{5}) picks out the unique worldline of a spinning test particle in the
gravitational field. That is, equations (\ref{1}), (\ref{2}) under (\ref{5}) do not admit
the oscillatory solutions. However, the question arises: is this worldline close, in the
certain sense, to the usual (non-helical) worldline of equations (\ref{1}), (\ref{2})
under condition (\ref{3}), for example, in a Schwarzschild field? It is simple to answer
this question when the relation
\begin{equation}\label{7}
m|u^\nu|\gg \left|u_\mu\frac {DS^{\nu\mu}}{ds}\right|
\end{equation}
takes place, because in this case condition (\ref{5}) practically coincides with
(\ref{3}). For example, these conditions are equivalent for post-Newtonian expansions
\cite{9}. However, {\it a priori} another situation is possible for the highly
relativistic spinning particle. Naturally, this situation must be investigated carefully.

Note that very condition (\ref{3}) was derived in some papers by different methods
\cite{10,11,12}, and we agree with the conclusion that this condition "... arises in a
natural fashion in the course of the derivation", \cite{11}, p.~112. That is, condition
(\ref{3}) is necessary, though often, with high accuracy, it can be substituted by
condition (\ref{5}).

We stress: the existence of the superfluous (oscillatory) solutions of equations
(\ref{1}), (\ref{2}) under condition (\ref{3}) is not a reason to ignore this condition.
The point of importance is that just among all solutions of equations
(\ref{1})--(\ref{3}) the single solution describing the motion of the particle's proper
center of mass can be found. Obviously, it is necessary to know how this solution can be
identified among others.

In the focus of this paper is just the initial Mathisson equations (\ref{1})--(\ref{3}).
Our purpose is to investigate non-oscillatory solutions describing highly relativistic
motions of the spinning particle in a Schwarzschild field.

Note that the information on all possible types of motions of the spinning test particles
in the gravitational fields is important for astrophysics, for more fine investigations
of the gravitational collapse and other astrophysical phenomena.

This paper is organized as follows. The known integrals of the strict Mathisson equations
(\ref{1})--(\ref{3}) in a Schwarzschild field, the energy and angular momentum, are used
in section 2 for reducing the order of differentiation in these equations. The possible
procedure of finding the values of the energy and momentum parameters, which correspond
just to the motions of the proper center of mass, is discussed in section 3. This
procedure is realized in section 4 for the motions close to highly relativistic
equatorial circular orbits in a Schwarzschild field. In section 5 the illustrations of
computer calculation are presented. We conclude in section 6.

\section{Mathisson equations for equatorial motions\\
in a Schwarzschild field}

Let us consider equations (\ref{1})--(\ref{3}) for Schwarzschild's metric in the standard
coordinates $x^1=r,\quad x^2=\theta,\quad x^3=\varphi,\quad x^4=t$ for the equatorial
motions of a spinning particle with spin orthogonal to the motion plane $\theta=\pi/2$.
Then the non-zero components of the metric tensor $g_{\mu\nu}$ are:
\begin{equation}\label{8}
g_{11}=-\left(1-\frac{2M}{r}\right)^{-1},\quad
g_{22}=g_{33}=-r^2,\quad g_{44}=1-\frac{2M}{r},
\end{equation}
where $M$ is the Schwarzschild mass. Due to the symmetry of Schwarzschild's metric
equations (\ref{1}), (\ref{2}) have the integrals of the energy $E$ and the angular
momentum $L$ which for the equatorial motions can be written as \cite{13}
$$
E=mu_4+g_{44} u_\mu \frac {DS^{4\mu}}{ds}+\frac12 g_{44,1}S^{14},
$$
\begin{equation}\label{9}
L=-mu_3-g_{33} u_\mu \frac {DS^{3\mu}}{ds}-\frac12 g_{33,1}S^{13}.
\end{equation}

For the equatorial motions with spin orthogonal to the motion plane $\theta=\pi/2$
equations (\ref{2}) can be solved separately from (\ref{1}). Indeed, taking into account
relations (\ref{3}) and (\ref{8}) it is not difficult to obtain from (\ref{2}) all
non-zero components $S^{\mu\nu}:$
$$
S^{13}=-S^{31}= -\frac{u_4 S_0}{r}, \quad S^{14}=-S^{41}=\frac{u_3 S_0}{r},
$$
\begin{equation}
\label{10} S^{34}=-S^{43}=-\frac{u_1 S_0}{r},
\end{equation}
where $S_0$ is the known constant of spin \cite{10}
\begin{equation}\label{11}
S_0^2=\frac12S_{\mu\nu}S^{\mu\nu}.
\end{equation}
(We stress that expressions (\ref{10}) satisfy all equations of set (\ref{2}).)

Now we shall consider equation (\ref{1}). It is known that these equations under
condition (\ref{3}) contain the second order derivatives $u^{\mu}$ with respect to $s$.
However, in our case of Schwarzschild's metric, due to the integral $E$ and $L$, it is
possible to reduce the order of differentiation by the standard procedure of the
differential equations theory. Using (\ref{8})--(\ref{10}) and the relation $u_\mu
u^\mu=1$ after direct calculations (which are quite simple but rather lengthly) we get
from equation (\ref{1}) the two non-trivial equations for $r(s)$ and $\varphi(s)$:
$$
\ddot r=\frac{\dot r^2}{r}+2r\left(1-\frac{3M}{r}\right)\dot \varphi^2-\frac{rE}{S_0}\dot
\varphi
$$
\begin{equation}\label{12}
 +\frac{1}{r}\left(1-\frac{3M}{r}\right)+\frac{L}{rS_0}\left[\dot
 r^2+\left(1-\frac{2M}{r}\right)(1+r^2\dot \varphi^2)\right]^{1/2},
\end{equation}
\begin{equation}\label{13}
\ddot \varphi=-\frac{\dot r\dot \varphi}{r}+r\left(1-\frac{3M}{r}\right)\frac{\dot
\varphi}{\dot r}+\frac{m+L\dot \varphi}{rS_0\dot r}\left[\dot
 r^2+\left(1-\frac{2M}{r}\right)(1+r^2\dot
 \varphi^2)\right]^{1/2},
\end{equation}
where a dot denote the usual derivatives with respect to $s$. (Two other equations of set
(\ref{1}) are satisfied identically.) So, equations (\ref{12}), (\ref{13}) do not contain
the third coordinate derivatives. However, in these equations the quantities $E$ and $L$
are present as the parameters which are not determined by the initial values of $r$,
$\varphi$, $\dot r\equiv u^1$, $\dot \varphi\equiv u^3$ only. (According to (\ref{9}),
for the determination of $E$ and $L$ the second coordinate derivatives must be given as
well.) That is, equations (\ref{12}), (\ref{13}), as well as the initial Mathisson
equations (\ref{1}), describe the motions both of the proper center of mass and the
non-proper centers.

In other words, equations (\ref{12}), (\ref{13}) contain the non-oscillatory and
oscillatory solutions. The question of importance is: which values $E$ and $L$ correspond
just to the proper center of mass at the arbitrary initial values of $r$, $\varphi$,
$\dot r$, $\dot \varphi$ for a spinning test particle? It is easy to answer this question
when the motion of such a particle is close to the geodesic motion: then approximately
$E=mu_4$ and $L=-mu_3$, i.e., we write the relations for the geodesic motion. However, we
have not any proof that the worldline of a spinning particle is close to the
corresponding geodesic worldline for all physically admitted initial values of $r$,
$\varphi$, $\dot r$, $\dot \varphi$. On the contrary, our results of investigations of
the gravitational spin-orbit interaction in a Schwarzschild field \cite{14,15,16} show
that just highly relativistic motions of a spinning particle must be studied carefully.
Therefore, in the next section we shall consider the procedure of choosing the values $E$
and $L$ for the proper center of mass for highly relativistic motions.

For further calculations it is convenient to write equations
(\ref{12}), (\ref{13}) in terms of the non-dimensional quantities
\begin{equation}\label{14}
\tau\equiv\frac{s}{M},\quad Y\equiv\frac{dr}{ds},\quad Z\equiv M\frac{d\varphi}{ds},\quad
\rho\equiv\frac{r}{M},\quad \varepsilon\equiv \frac{S_0}{mM}.
\end{equation}
(In the following we shall put $\varepsilon>0$, without any loss
in generality.) Then according to equations (\ref{12}), (\ref{13})
we have the set of the first-order differential equations
$$
\frac{dY}{d\tau}=\frac{Y^2}{\rho}+\rho\left(1-\frac{3}{\rho}\right)\left(2Z^2+\frac{1}{\rho^2}\right)-\mu
Z\rho
$$
\begin{equation}\label{15}
+\frac{\nu}{\rho}\left[Y^2+\left(1-\frac{2}{\rho}\right)(1+Z^2\rho^2)\right]^{1/2},
\end{equation}
$$
\frac{dZ}{d\tau}=-\frac{YZ}{\rho}+\rho\frac{Z^2+1/\rho^2}{Y}\left(Z-\frac{3Z}{\rho}-\mu\right)
$$
\begin{equation}\label{16}
+ \frac{1}{\rho Y}\left(\frac{1}{\varepsilon} +\nu
Z\right)\left[Y^2+\left(1-\frac{2}{\rho}\right)(1+Z^2\rho^2)\right]^{1/2},
\end{equation}
\begin{equation}\label{17}
\frac{d\rho}{d\tau}=Y,
\end{equation}
\begin{equation}\label{18}
\frac{d\varphi}{d\tau}=Z,
\end{equation}
where
\begin{equation}\label{19}
\mu\equiv\frac{ME}{S_0},\qquad \nu\equiv\frac{L}{S_0}.
\end{equation}
For correct studying the physical conclusions of equations (\ref{15})--(\ref{18}), it is
necessary to take into account the Wald condition for a test spinning particle \cite{17},
therefore we put
\begin{equation}\label{20}
\varepsilon\equiv \frac{S_0}{mM}\ll 1.
\end{equation}
Because in the following we shall compare some solutions of equations
(\ref{15})--(\ref{18}) with solutions of the geodesic equations in Schwarzschild's
metric, it is useful to write here the last equations for the equatorial motions in the
coordinates $r, \varphi$
\begin{equation}\label{21}
\ddot r=\dot\varphi^2 r\left(1-\frac{3M}{r}\right)-\frac{M}{r^2},
\end{equation}
\begin{equation}\label{22}
\ddot \varphi=-\frac{2}{r}\dot r\dot\varphi.
\end{equation}
Using notation (\ref{14}) we rewrite equations (\ref{21}), (\ref{22}) as the set of the
first-order differential equations
\begin{equation}\label{23}
\frac{dY}{d\tau}=Z^2\rho\left(1-\frac{3}{\rho}\right)-\frac{1}{\rho^2},
\end{equation}
\begin{equation}\label{24}
\frac{dZ}{d\tau}=-2\frac{YZ}{\rho},
\end{equation}
plus two equations which coincide with (\ref{17}), (\ref{18}). Equations (\ref{15}),
(\ref{16}) and (\ref{23}), (\ref{24}) correspondingly are essentially different. As well
as equations (\ref{12}), (\ref{13}), the first two equations of set
(\ref{15})--(\ref{18}) contain the constants of the energy and angular momentum of a
spinning particle. According to discussion above, different values of the parameters
$\mu$ and $\nu$ in (\ref{15}), (\ref{16}) at the fixed values of $Y, Z, \rho$ correspond
to different centers of mass, namely to the single proper center of mass and to the set
of non-proper centers.

\section{Possible way of finding the parameters $\mu$ and $\nu$ for the proper center of mass}

The right-hand sides of equations (\ref{15}), (\ref{16}) are too complicated for
investigations in general case. However, {\it a priori} we cannot  exclude the
possibility to show the essential difference between the motions of the proper and
non-proper centers of mass during the short time interval after the beginning of the
corresponding motions, namely, when the displacement of the values $Y, Z, \rho, \varphi$
from their initial values $Y_0, Z_0, \rho_0, \varphi_0$ are considered in the linear
approximations in the quantities
\begin{equation}
\label{27} \xi _{1} \equiv \frac{Y - Y_{0}}{Y_{0}}, \quad \xi _{2}
\equiv \frac{Z - Z_{0}}{Z_{0}}, \quad \xi _{3} \equiv \frac{\rho -
\rho _{0}}{\rho _{0}}.
\end{equation}
(In (\ref{27}) we do not write the displacement $\varphi-\varphi_0$ because the
right-hand sides of equations (\ref{15})--(\ref{18}) do not depend on $\varphi$ and
equation (\ref{18}) is trivial, i.e., according to (\ref{18}) the value $\varphi$ is
determined by simple integration of $Z$.) Let us check this possibility, i.e., consider
equations (\ref{15})--(\ref{17}) in the linear in $\xi$ approximation. Then by direct
calculations we obtain
$$
\frac{d\xi _{1}}{d\tau} = \left( a_{10} + a_{11} \nu + a_{12} \mu
\right)\xi _{1} + \left( {a_{20} + a_{21} \nu + a_{22} \mu}
\right)\xi _{2}
$$
\begin{equation}
\label{28}
+ \left( a_{30} + a_{31} \nu + a_{32} \mu \right)\xi
_{3} + a_{00} + a_{01} \nu + a_{02} \mu ,
\end{equation}
$$
\frac{d\xi _{2}}{d\tau} = \left( b_{10} + b_{11} \nu + b_{12} \mu
+ b_{13} \frac{1}{\varepsilon} \right)\xi _{1} + \left( b_{20} +
b_{21} \nu + b_{22} \mu + b_{23} \frac{1}{\varepsilon} \right)\xi
_{2}
$$
\begin{equation}
\label{29} + \left( b_{30} + b_{31} \nu + b_{32} \mu + b_{33}
\frac{1}{\varepsilon} \right)\xi _{1} + b_{00} + b_{01} \nu +
b_{02} \mu + b_{03} \frac{1}{\varepsilon},
\end{equation}
\begin{equation}
\label{30} \frac{d\xi _{3}} {d\tau} = c_{10} \xi _{1} + c_{00},
\end{equation}
where the coefficients $a, b, c$ with the corresponding indexes are expressed through
$Y_0, Z_0, \rho_0.$ We shall use the expressions of these coefficients in the
approximation
\begin{equation}\label{31}
Z_0^2\rho_0^2\gg 1, \quad Y_0^2\ll Z_0^2\rho_0^2.
\end{equation}
According to notation (\ref{14}),  relations (\ref{31}) mean that the tangential
component of the particle's initial velocity is highly relativistic, and, in addition,
that the tangential component is much greater than the radial component. First, it
corresponds with our aim to investigate  just highly relativistic motions. Second,
because the deviation of a spinning particle is caused by the gravitational spin-orbit
interaction, most clearly this deviation can be shown in the case when the tangential
velocity is dominant (in particular, for the circular or closer to the circular orbits
\cite{16}).

According to the theory of differential equations, the general solution of linear
equations (\ref{28})--(\ref{30}) is determined by the combination of $e^{\lambda_i\tau}$
(i=1,2,3), where $\lambda_i$ are the solutions of the third-order algebraic equation
\begin{equation}
\label{33} \lambda ^{3} + C_{2} \lambda ^{2} + C_{1} \lambda + C_{0} = 0.
\end{equation}
Here the coefficients $C_j$ ($j=0,1,2$) can be expressed through $a, b, c$ and depend
both on $\rho_0, Y_0, Z_0, \varepsilon$ and on the parameters $\mu, \nu$. Our task is to
find such concrete values $\mu, \nu$ which at the fixed $\rho_0, Y_0, Z_0, \varepsilon$
determine just the motion of the proper center of mass. We begin by analyzing the
expressions $C_j$ for the simple case when the worldline of a spinning particle is close
to the corresponding geodesic worldline.

\subsection{Expressions $C_j$ for quasi-geodesic motions}

If the parameter $\varepsilon$ in (\ref{20}) is sufficiently small, for any fixed values
$\rho_0, Y_0, Z_0$ the motion of the proper center of mass is close to the geodesic
motion. Then we can write approximately $E=mu_4, \quad L=-mu_3$ (see expressions
(\ref{9})), and according to (\ref{14}), (\ref{19}) we have
\begin{equation}
\label{41} \mu = {\frac{{1}}{{\varepsilon} }}{\left| {Z_{0}} \right|}\rho _{0} \left( {1
- {\frac{{2}}{{\rho _{0}} }}} \right)^{1/2} , \quad \nu = {\frac{{1}}{{\varepsilon}
}}\rho _{0}^{2} Z_{0}.
\end{equation}
Let us consider expressions $C_j$ at $\mu, \nu$ from (\ref{41}).  It is easy to show that
all these expressions  have a common feature: all greatest terms with the large value
$1/\varepsilon$ are cancelled. It means that the corresponding largest terms with
$1/\varepsilon$ are absent in the expressions $\lambda_j$ as well. So, just the values
$\mu, \nu$ for the proper center of mass give the minimum values of $\lambda_j$. We
stress that similar situation takes place for motions of the proper center of mass in the
Minkowski spacetime: for the proper center $\lambda_j=0$ (it corresponds to the
straightforward motions), and for the non-proper centers $\lambda_j$ are proportional to
$M/S_0$ (the oscillatory motions).

\subsection{Expressions $C_j$ for the highly relativistic circular motion
with $r=3M$}

In \cite{16} we have considered the case of the circular motion of the proper center of
mass in a Schwarzschild field with $r=3M$. By the notation (\ref{14}) in this case we
write
\begin{equation}
\label{41a} \rho = 3, \quad Y = 0, \quad Z = - \frac{3^{-3/4}}{\sqrt{\varepsilon}}.
\end{equation}
For values (\ref{19}), from (\ref{9}) we obtain
\begin{equation}
\label{42} \mu = 3^{-1/4}\varepsilon^{-3/2}, \quad \nu = - 3^{5/4}\varepsilon^{-3/2}.
\end{equation}
Let us estimate the values $C_j$. Taking into account (\ref{42}) it is easy to check
that, as well as in the previous case, in all expressions $C_j$ the largest terms with
$1/\varepsilon$ are cancelled.

It is naturally to suppose that the similar  feature takes place not only in the two
partial cases above. Therefore, below we shell check do the criterion of excluding the
largest terms with $1/\varepsilon$ in the expressions $C_j$ can be used for finding the
values $\mu$ and $\nu$ which pick out just the motion of the proper center of mass.

\section{Expressions $\mu$ and $\nu$ for the motions of the proper center of mass
at $2<\rho_0 <3$ under condition (\ref{31})}

As we pointed out in \cite{16} equations (\ref{1})--(\ref{3}) admit in a Schwarzschild
field the solutions describing the circular equatorial orbits in the region $2M<r<3M$. By
notation (\ref{14}) for these orbits we have
\begin{equation}
\label{43} 2 < \rho < 3, \quad Y = 0, \quad Z = - {\frac{{1}}{{\rho} }}\left( {1 -
{\frac{{2}}{{\rho} }}} \right)^{1/4}{\left| {1 - {\frac{{3}}{{\rho} }}}
\right|}^{-1/2}\varepsilon ^{-1/2}.
\end{equation}
The expression $Z$ from (\ref{43}) is valid beyond the small neighborhood of the value
$\rho=3$. (This neighborhood was considered in \cite{16}.) At expressions (\ref{43}), it
follows from (\ref{9}) that
\begin{equation}
\label{44} \mu \sim \varepsilon ^{-1/2}, \quad \nu \sim \varepsilon ^{-1/2}.
\end{equation}
It is interesting to consider the non-circular orbits which deviate from (\ref{43}) due
to $Y_0\ne 0$ under condition (\ref{31}). Let us consider the expressions $C_j$ in the
case
\begin{equation}
\label{45} 2 < \rho _{0} < 3, \quad 1 < < Y_{0}^{2} < < Z_{0}^{2} \rho _0^2, \quad Z_{0}
= k\varepsilon ^{-1/2},
\end{equation}
where according to (\ref{43}), (\ref{44})
\begin{equation}
\label{46} k = - \frac{1}{\rho_0} \left( 1 - \frac{2}{\rho_0} \right)^{1/4}\left| 1 -
\frac{3}{\rho_0} \right|^{-1/2},
\end{equation}
\begin{equation}
\label{47} \mu = k_1 \varepsilon ^{-1/2}, \quad \nu = k_2\varepsilon ^{-1/2} ,
\end{equation}
and $k_1$, $k_2$ are some coefficients which do not depend on $\varepsilon$. Our task is
to find such values $k_1$, $k_2$ which ensure excluding the largest terms with
$1/\varepsilon$ in the expressions $C_j$.

Using $\mu$,  $\nu$ from (\ref{46})  we obtain the conditions under which the largest
terms with $1/\varepsilon$ in the expressions $C_j$ are cancelled:
\begin{equation}
\label{48} 3k^2(\rho _{0} - 3) - 2\rho _{0}k k_{1} - 2k\left( 1 - \frac{2}{\rho _0}
\right)^{1/2}k_{2} - \left( 1 - \frac{2}{\rho _{0}} \right)^{1/2} = 0,
\end{equation}
\begin{equation}
\label{49}
 - k^{2}(\rho _{0} - 3) + \rho _{0} kk_{1} + kk_{2} \left( {1 -
{\frac{{2}}{{\rho _{0}} }}} \right)^{1/2} + \left( 1 - \frac{2}{\rho _{0}} \right)^{1/2}
= 0,
\end{equation}
\begin{equation}
\label{50} k^{2}\rho _{0} - \rho _{0} kk_{1} - kk_{2}\frac{1}{\rho _0}\left( 1 -
\frac{2}{\rho _0} \right)^{-1/2} - \frac{1}{\rho _0}\left( 1 - \frac{2}{\rho _0}
\right)^{-1/2} = 0.
\end{equation}
It is easy to check that at the value $k$ from (\ref{46}) among three linear in $k_1$,
$k_2$ algebraic equations (\ref{48})--(\ref{50}) there are only two independent
equations. Then the solution of (\ref{48})--(\ref{50}) is:
\begin{equation}
\label{51} k_1 = - \frac{1}{\sqrt {\rho _0}}\left( 1 - \frac{2}{\rho _0}
\right)^{1/4}\left| 1 - \frac{3}{\rho _0} \right|^{-3/2}\left( 1 - \frac{3}{\rho _0} +
\frac{3}{\rho _0^2} \right),
\end{equation}
\begin{equation}
\label{52} k_2 = \sqrt{\rho_0}\left( 1 - \frac{2}{\rho _0} \right)^{-1/4}\left| 1 -
\frac{3}{\rho _0} \right|^{-3/2}\left( 1 - \frac{9}{\rho _0} + \frac{15}{\rho _0^2}
\right).
\end{equation}
Relations (\ref{51}), (\ref{52}) can be used in computer integration of equations
(\ref{15})--(\ref{18}) under conditions (\ref{45}).

\section{Examples of computer integration of equations (\ref{15})--(\ref{18})}

Graphs of $\rho(\tau)$ and $\rho(\varphi)$ according to equations (\ref{15})--(\ref{18})
under relations (\ref{51}), (\ref{52}) are shown by the thick lines in figures 1 and 2
correspondingly. For comparison, the thin lines in the same figures show graphs of
$\rho(\tau)$ and $\rho(\varphi)$ according to geodesic equations (\ref{23})--(\ref{24}).
If relations (\ref{51}), (\ref{52}) are violated, computer integration shows the
oscillatory solutions. Similar situation takes place for orbits with $\rho_0=3$. An
example of the oscillatory solution is presented in figure 3.

\vspace{2mm}
\begin{figure}[h]

\end{figure}
{\small Fig. 1. Graph of $\rho(\tau)$ according to equations (\ref{15})--(\ref{18}) under
relations (\ref{51}), (\ref{52}) at $\varepsilon=10^{-6}$, $\rho_0=2.5$, $Y_0=0.3$ and
$Z_0$ determined by (\ref{45}), (\ref{46}) (thick line). Graph of $\rho(\tau)$ according
to geodesic equations (\ref{23})--(\ref{24}) at the same $\rho_0$, $Y_0$,  $Z_0$ (thin
line). The horizontal line $\rho=2$ corresponds to the horizon surface. }

\vspace{2mm}
\begin{figure}[h]

\end{figure}
{\small Fig. 2. Graph of $\rho(\varphi)$ in the polar coordinates according to equations
(\ref{15})--(\ref{18}) (thick line) and (\ref{23})--(\ref{24}) (thin line).
 The values $\rho_0$, $Y_0$, $Z_0$ are the same as in Fig.~1. The circle $\rho=2$ corresponds to the horizon
 surface.}

\vspace{2mm}
\begin{figure}[h]

\end{figure}
{\small Fig. 3. An example of the oscillatory solution for $\rho(\tau)$ according to
equations (\ref{15})--(\ref{18}) at $\varepsilon=10^{-4}$, $\rho_0=3$, $Y_0=2.5\times
10^{-3}$ and $Z_0$ determined by (\ref{41}) when relations (\ref{42}) are slightly
violated.}

\section{Summary}

The computer integration of equations (\ref{15})--(\ref{18}) at relations (\ref{51}),
(\ref{52}) shows that these relations are suitable for choosing solutions, which describe
the motions of the proper center of mass. As a result, according to figures 1 and 2, we
conclude that (under those conditions indicated in the captions at these figures) the
force of the interaction of the particle's spin with the gravitational field acts as the
repulsive one. Due to this force the spinning particle falls on the horizon surface
during longer time as compared to the corresponding particle without spin (figure 1).
Moreover, according to figure 2,  considerable space separation of the corresponding
spinning and non-spinning particles takes place within a short time, i.e., within the
time of the particle's fall on the Schwarzschild horizon.

There is also a possibility of generalizing the above onto case of a Kerr field.

R.P. thanks Professors A. Trautman and E. Malec for useful discussions and hospitality in
Warsaw and Cracow.

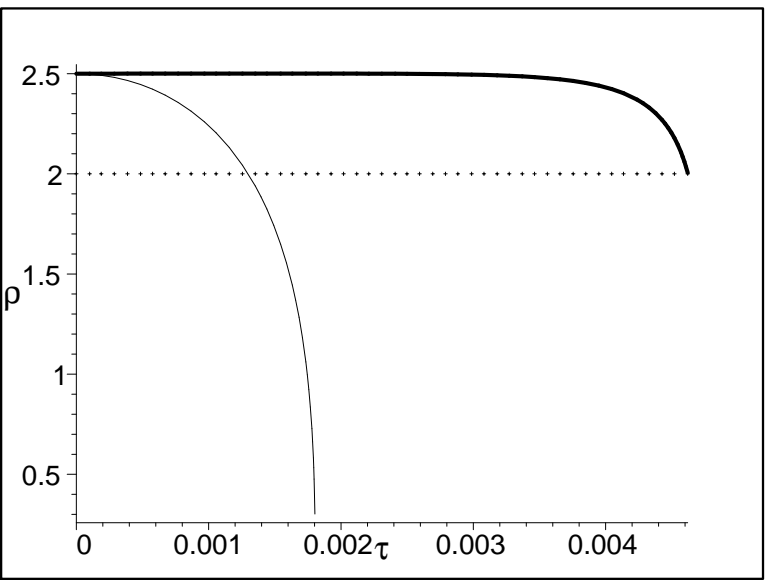
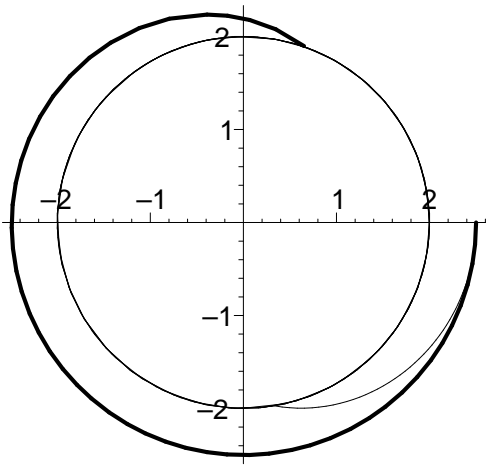
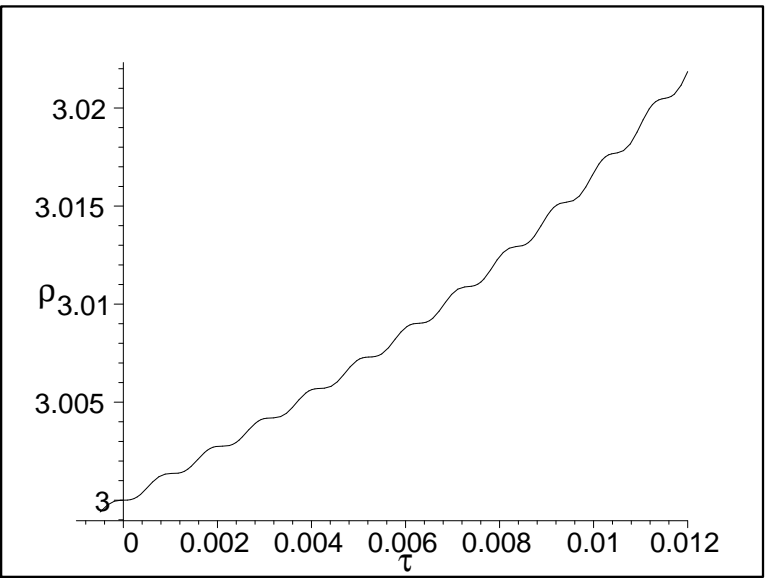
\end{document}